# Enhancing Distributed Authorization with Lagrange Interpolation and Attribute-Based Encryption

Keshav Sinha [1], Sumitra [1], Richa Kumari [1], Akashdeep Bhardwaj [1] and Shawon Rahman [2]

[1] Centre for Cybersecurity, School of Computer Science, UPES, Dehradun, India, 248007
[2] Dept. of Computer Science, University of Hawaii-Hilo, Hilo, HI 96720, USA

## ABSTRACT

*In today's security landscape, every user wants to access largeamounts of data with confidentiality and authorization. To maintain confidentiality, various researchers have proposed several techniques; however, to access secure data, researchers use access control lists to grant authentication and provide authorization. The above several steps will increase the server's computation overhead and response time. To copewith these two problems, we proposedmulti-party execution on the server. In this paper, we introduce two different approaches. The first approach is encryption, utilizing the Involution Function-Based StreamCipherto encrypt the file data. The second approach is key distribution, using the Shamir secret-sharing scheme to divide and distribute the symmetric key to every user. The decryption process required key reconstruction, which used second-order Lagrange interpolation to reconstruct the secret keys from the hidden points. The process will reduce the server's computational overhead. The results are evaluated based on the encryption and decryption time, throughput, computational overhead, and security analysis. In the future, the proposed mechanism will be used to share large-scale, secure data within the organization.*

## KEYWORDS

*Involution Function, Stream Cipher, Distributed Authorization, Attribute-Based Encryption, LCG, XOR, Access Control List, Lagrange Interpolation, Shamir's Secret Sharing*

## 1. INTRODUCTION

Organizations have been struggling with the challenge of managing and storing a vast amount of data in recent times. Consequently, most data owners have turned to cloud servers, which are typically supported by predefined access control policies (ACP), to store their data. Cloud computing can deliver cheaper and more efficient resources to users through three main service models: Platform as a Service (PaaS), Software as a Service (SaaS), and Infrastructure as a Service (IaaS). But since the data are under the management of Cloud Service Providers (CSPs), they become exposed to unauthorized access and hacking activities [1]. By moving data to the cloud, organizations can save on local management costs, but they also give up some control over privacy and security. Regular encryption techniques can still be used to secure data before it is uploaded to the cloud; however, these methods suffer from issues such as complex key management and limited storage space [2].

The proposed architecture combines CP-ABE [3] with ACLs and a re-encryption mechanism to enable the transfer of ciphertext sharing. The methodology seeks to address challenges associated with three primary facets: key management, authorization, and privacy protection. The authors propose employing symmetric meta-encryption and Shamir's Secret Sharing for key generation and distribution [4], with policy modifications dynamically reflected in data confidentiality updates. Lagrange interpolation is used to reassemble the decryption keys, whereas a lightweight stream cipher using an involution function is chosen for encryption. The proposal reduces computation and storage overhead while maintaining strong access control[5][6].This approach prevents the leakage of private data [6]. The key features of the proposed system are:

• Involution Function-Based Stream Cipher has been used to create the ciphertext.
• Sharing ciphertext using a simple access control scheme.
• Lagrange interpolation has been used to reconstruct decryption keys
• Reduce the overhead of key sharing using multilinear mapping.
• Performance evaluation is based on the encryption time and storage overhead.

The rest of this paper is structured as follows: Section 2 discusses related works; Section 3 describes the proposed system architecture; Section 4 presents results and performance evaluation; and Section 5 concludes with key findings and future scope.

## 2. RELATED WORK

Data owners opt to store their data with cloud service providers (CSPs), entrusting them with its preservation. However, rather than relying solely on the CSP, the data is encrypted before transmission to the cloud, ensuring confidentiality.

### 2.1. Access Control on Encrypted Data

Access control is one of the key factors that determine how safe and private encrypted data is, especially if the data is sensitive and stored or shared in the cloud. Usually, the data are encrypted, and the decryption keys are given only to authorized users. When trust levels among users or systems change, the keys may need to be updated to maintain security.In a conventional symmetric cryptography scenario, data owners primarily use Access Control Lists (ACLs) to categorize files, then encrypt each category with a different key. This guarantees confidentiality; however, keys scale linearly with the number of data groups, so key management becomes twice as complex, and storage capacity doubles [7].

The system's performance can be improved by using a combined method that implements both symmetric and public-key encryption. Such systems encrypt data with a symmetric key, and then each user's public key is used to encrypt the symmetric key. This provides a secure one-time distribution of the key and individual decryption rights, but increases computational cost with the number of users [8].

Attribute-Based Encryption (ABE) is a more adaptable solution that grants access based on user attributes rather than fixed identities. ABE is a system in which only users with attributes that satisfy the encryption policy can access the data. There are two types of ABE Key-Policy (KP-ABE) and Ciphertext-Policy (CP-ABE). CP-ABE is widely used in the cloud; however, its high computational requirements remain a problem [9]. Some studies have also reported the adoption of reputation-based access strategies as a recent topic. One example is Shamir's Secret Sharing, which provides confidentiality in multi-cloud environments by verifying secret shares using hash signatures [10]. Similarly, Sinha et al. [11] compared ABE models with the DAC-MACS framework, which uses multiple authorities and certificate validation to prevent collusion and enhance key revocation efficiency.

## 2.2. User Revocation

Revocation is a crucial factor in maintaining well-controlled, safe access in cloud environments. Once the data owner decides to withdraw user access, i.e., members who leave a group or become untrust worthy, the revoked users may still have old decryption keys. To counteract these unauthorized actions, the data owner should re-encrypt the data using new keys and share them with the legitimate users [12]. Nevertheless, the problem of handling such a process arises in Attribute-Based Encryption (ABE), which, in turn, leads to frequent re-encryption and key redistribution, thereby drastically increasing computational overhead and management complexity [13].

It is viable to additionally use a semi-trusted third party to take over the re-encryption work from you. Proxy Re-Encryption (PRE) authorizes a proxy to alter the encrypted data that one user can decrypt so that another user can decrypt the same data, without the proxy learning the plaintext. In the case of ABE integration, PRE supports scalable user and attribute revocation by delegating partial encryption rights to the Cloud Service Provider (CSP). However, it is still necessary that the data owner be present to issue PRE keys, which can delay revocation and pose security risks [14][15][19][20].

We have a model that features a Reputation Center, which represents the data owner in this matter. It actively engages in re-encryption tasks and ensures that revoked users cannot re-access the system. By doing this, only the data encryption key is re-encrypted rather than the whole dataset, thus the computational load on the CSP is very low. This reputation-driven approach not only helps become more efficient but also better manages trust in distributed environments [16].Though some reputation-based schemes have been proposed, most fail to regulate access control based on user credibility strictly. The use of reputation mechanisms in conjunction with appropriate cryptographic schemes may lead to the next innovation in the security and privacy of large-scale cloud systems [17][19].

## 3. RESEARCH METHODOLOGY

Figure 1 presents the framework for an organizational environment in which data is securely shared among multiple users. The four key participants are (i) Data Owner, (ii) Cloud Storage, (iii) Data User, and (iv) Organization Server [11].

- Data Owner: A registered user who creates files and defines an Access Control List (ACL) for sharing within the organization. The attributes include User_Id ($U_i$), User_Type ($UT_i$), and User_Credentials ($UC_i$).
- Cloud Storage: A semi-trusted platform that stores encrypted data and backup ACLs. It follows access lists to manage authorized data operations.
- Data User: A verified user listed in the authorization database. Access is granted through authorization points linked to valid credentials.
- Organization Server: Serves as the intermediary for key distribution, encryption, and decryption. It securely stores ACL records, assigns or revokes access, and manages user credentials.

The comprehensive process for file encryption, authorization key distribution, and secure file storage is outlined as follows:

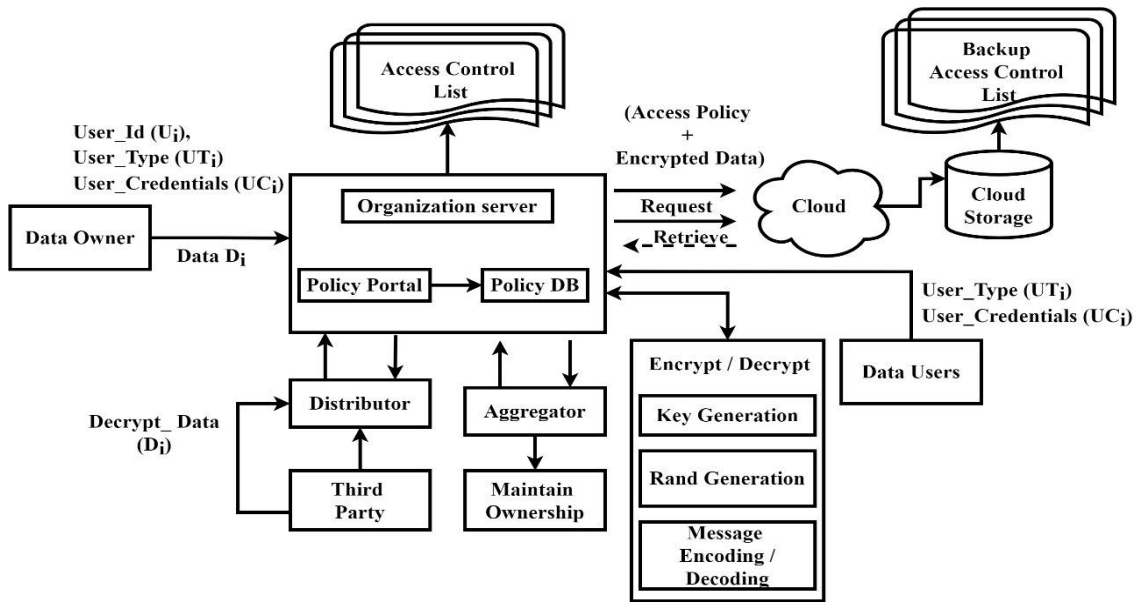

Figure 1. Distributed Access Control Scheme Model

**Procedures for Distribution of Authorization Keys and Encryption[11]**
The organization server, which is in charge of storing user credentials, is registered with the data owner.
File submission takes place on the organization's server.
A random bit [N1, N2] is generated and then XORed.
Firstly, the XORed random bit is XORed again with the original file. Then, functional-based encryption is applied to the resulting XOR data.
By using the Shamir Secret Sharing technique and the file owner's attributes, the server creates multiparty authorization points.
The authorization points are encrypted and handed over to the decrypting party.

### 3.1. System Functionality

This section describes the proposed implementation of the Distributed Authorization Scheme. The prime functions are (i) XOR Operation, (ii) Key Management, and (iii) Access Control Policy.

### 3.1.1. XOR Operation

The generation of a random bit through the application of the XOR operation using the Linear Congruential Generator (LCG) can be denoted as follows: Rand(•) function, exhibiting a probabilistic nature, produces a fixed random bit [Nk → N1], while the Rep(•) function generates a repeated random string [Nk → N2]. The resulting random bits [N1, N2] are XORed and again XOR with the original message at ΔT time intervals.

### 3.1.2. Key Management

For safe data encryption and user sharing, the organization's server uses a hybrid key structure. It maintains symmetric keys that are kept with the ciphertext and connected to file characteristics. Every file in the directory structure has a distinct key through a hierarchical key arrangement.

The Data Owner Secret Key and the Multiparty Authorization Key are the two primary keys used in the suggested model.

### 3.1.3. Access Control Policy

The organization server uses the policy portal and Policy DB to track user access lists for every file. For proper file-access permissions and the corresponding authorization key, the participant's credentials are required.

### 3.2. Involution Function-Based Stream Cipher -Based File Encryption

The Stream Based on Involution Functions Data files are encrypted on the server using a cipher before being uploaded to the cloud. It encrypts and decrypts data using deterministic procedures. An involutory function based on a bijective mapping is used for key building in order to reduce susceptibility from low-complexity systems. Since the members of sets A and B in this function are equal, the inverse mapping f: G → G´ satisfies f(AB) = f(B) f(A), indicating that the function "f" is an anti-homomorphism [11]. The involution function is defined by Eq. 1.

$$f(x) = X \qquad (1)$$

The reverse is written as Eq. 2.

$$f(f(x)) = X \qquad (2)$$

The inverse involution function is also known as Anti-involution (or Anti-homomorphism). A similar mechanism is applied to the encryption process as shown in Eq. 3, where a symmetric key is used for encryption and decryption, as shown in Eq. 4.

$$\text{Encryption} = D_{encrypted_i} = E_k(D_i) \qquad (3)$$

$$\text{Decryption} = D_i = D_k(D_{encrypted_i}) \qquad (4)$$

Here, $D_i$ = data, $E_k$ = encryption key, $D_K$ decryption key. The encryption method proposed in this system utilizes a stream cipher to process input data. The involution function is designed to operate on data streams. This involution function-based stream cipher is designed to be compatible with various electronic devices. As previously mentioned, the involution function possesses a unique property: it can restore the original content when applied again. The proposed Involution Function-Based Stream Cipher for encryption is defined by Eq. 5.

$$\text{Encryption} = f(x) = \left(a - x^{\frac{1}{n}}\right)^n \qquad (5)$$

Now, we compute the anti-evolution function $f(f(x))$, which we get by Eq. 6.

$$\text{Decryption} = f(f(x)) = \left(a - \left(\left(a - x^{\frac{1}{n}}\right)^n\right)^{\frac{1}{n}}\right)^n = x \qquad (6)$$

Here, 'a and 'n' are the variables and 'x' is the plaintext, and $f(x)$ is the ciphertext, where $a, x \in I^+$. To fit the formula with the cryptographic aspect, we set $a$' = Secret key and 'n' = Randomized power variable. The key storage is based on the data file, where one key is linked to one file. If the user changes the file name, it means that the authorization of the file is also changed, as represented by Eq. 7.

$$FK_i = (a_i \oplus x_i) \qquad (7)$$

Where '$a_i$' and '$a'_i$' are the old and new secret keys, the '$x_i$' is the full name for the file '$i$'. The next section encrypts the Secret key '$a$' with the owner's attributes to provide authorization only for privileged users.

## 3.3. CP-ABE-based Multiparty Authorization Key Computation

This section explains the implementation of CP-ABE-based multiparty authorizationkey generation. Authorization keys are distributed when the data owner shares the ciphertext with defined access privileges, ensuring only authorized users can decrypt the data [11]. The scheme applies Shamir's Secret Sharing for key distribution using a 1-degree Lagrange interpolation function over the finite field $Zp$.

### 3.3.1. Shamir Secret Sharing

It is ideal for the (k, n) threshold scheme, which divides secret key 'S' into 'n' pieces of data [$S_1$, $S_2$, …, $S_n$]. Only if the adversary knows any 'k' or more '$S_i$' pieces, can 'S' be easily computable (Liu et al., 2024). The reconstruction of secret 'S' is the combination of 'k' pieces of the data. The general formula for the construction of polynomials is represented as Eq. 8.

$$f(x) = a_0 + a_1 x + a_2 x^2 + \cdots + a_{k-1} x^{k+1} \qquad (8)$$

Here, $a_0$ = secret key, and user attributes token ∈ {$a_1$, $a_2$, …, $a_{k-1}$}. The key distribution is present as points, where n users on the server have privileged access policies. The server creates the user attribute tokens sets [i = (1, 2, 3, …, n )] to retrieve [$i$, $f(i)$]. Now, every user has given the points (a non-zero integer input to the polynomial). For every given subset of 'k', we can obtain the '$a_0$' using the 2-degree Lagrange interpolation. The secret key construction function is defined over the finite field $Z_p$, given by Eq. 9.

$$G(X) = (a_0 + a_1 X + a_2 X^2) \bmod p \qquad (9)$$

The key distribution is based on the number of users present on the server with privileged access.

---
**Algorithm 1: Key Distribution**
**01:** Start
**02:** $D_X$ = number of privileged users
**03:** Point generation from polynomials: $D_{X-1}$ = (X, G(X))
**04:** Point calculation:
**05:** Y = (G(X))
**06:** $D_{X-1}$ = (X, Y), where [X = 1, 2, …, $D_X$]
**07:** End
---

Now, single-point sharing is performed for each participant (x, G(x)). Here, the points start from (1, G(1)) and not from (0, G(0)) because G(0) consists of secret key information. The distribution of $x_i$'s key is only to privileged users and is simplified by the Distributor and Aggregator. The points are present in the hash formed (0, $KC_i$). $KC_i$ is the file binding code of '$x_i$', calculated using Eq. 10.

$$KC_I = (Sec_{K_i}, f(x)) \bmod p \qquad (10)$$

The interpolation function of file '$x_i$' must pass at least three points to generate the equation. The Lagrange basis polynomials are used to generate the points, and the formulation is given as Eq. 11:

$$l_j(x) = \prod_{\substack{0<m\le k \\ m\ne j}} \frac{x-x_m}{x_j-x_m} \qquad (11)$$

The system assumption is that no two points '$x_j$' are the same, (m ≠ j) and $(x_j - x_m) \ne 0$. Consider an n-user on the server with the privileged access policy, the users having the instances set (i = 1, 2, 3, …, n). Eq. 8 is used to distribute the key between different users. All users have given the points (a non-zero integer input polynomial). Now, for every given subset of decryption key 'k', we can obtain '$a_0$' using the interpolation.

### 3.3.2. Key Generation with Parabolic Equation

The organization server generates multiparty authorization points by applying the Shamir Secret Sharing technique to the file owner's attributes. Suppose our secret key 'k = 1234'. We define the equation for the distribution of the key. Here, we use the parabolic equation to construct the distribution. It is defined as Eq. 12:

$$F(x) = a_0 + a_1 X + a_2 X^2 \qquad (12)$$

Here, '$a_1$' and '$a_2$' are the attributes of the owner, and '$a_0$' is the secret key. Now Eq. 13 represents the symmetric key and user attributes.

$$F(X) = 94X^2 + 166X + 1234 \qquad (13)$$

Then, the Number of Privileged Users ($D_X$) = 6, and $D_{X-1}$ = (X, F(X)) from the polynomial.

$$Y = F(X) = 1234 + 166X + 94X^2 \qquad (14)$$

$$D_{X-1} = (X, Y) \text{ where } [X = 1, 2, 3, 4, 5, \text{ and } 6] \qquad (15)$$

Now, solve Eq. 14 using Eq. 15, and the value of '$D0$' is presented in Table 1.

Table 1: Each Participant's Single Points

| $D_{X-1}$ | $F(X_n, Y_n)$ | Y |
|---|---|---|
| $D_0$ | $F(X_0,Y_0)$ | 1, 1494 |
| $D_1$ | $F(X_1,Y_1)$ | 2, 1942 |
| $D_2$ | $F(X_2,Y_2)$ | 3, 2578 |
| $D_3$ | $F(X_3,Y_3)$ | 4, 3402 |
| $D_4$ | $F(X_4,Y_4)$ | 5, 4414 |
| $D_5$ | $F(X_5,Y_5)$ | 6, 5614 |

Now, we give each participant their share of the authorization key (X, F(X)). Here, $D_{X-1}$ instead of $D_X$ is shared with the participant because points start from (1, F(1)), not from (0, F(0)), because F(0) is the secret key value. The organization server maintains the DB policy based on the participant decryption points.

### 3.4. Authorization Key Reconstruction

The key reconstruction is performed on the organization server. The key reconstruction requires a minimum of three points: (i) Organization Server Point, (ii) Owner point, and (iii) Receiver point to generate the parabolic equation[11]. The organization server does not store the owner points and contributions of all parties in the proposed scheme to mitigate Denial-of-Service (DoS) attacks. The decrypted points $(X, F(X))$ are used for the interpolation. Let us consider $F(X_1, Y_1)$ = Organization server point, $F(X_3, Y_3)$ = Owner point, and $F(X_4, Y_4)$ = Receiver Point, and the respective values are (2, 1942), (4, 3402), and (5, 4414) for the key reconstruction. The Lagrange basic polynomial is represented as Eq. 16.

$$l_j(x) = \prod_{\substack{0<m\leq k \\ m\neq j}} \frac{x-x_m}{x_j-x_m} => \frac{x-x_0}{x_j-x_0} \cdots \frac{x-x_{j-1}}{x_j-x_{j-1}} \times \frac{x-x_{j-1}}{x_j-x_{j+1}} \cdots \frac{x-x_k}{x_j-x_k} \quad (16)$$

Where $0 \leq j \leq k$, the assumption is that there are no two $x_j$ are the same, where $(m \neq j)$ and $(x_j - x_m) \neq 0$. Now,

$$l_0(x) = \frac{x-x_1}{x_0-x_1} \times \frac{x-x_2}{x_0-x_2} \quad (17)$$

$$l_1(x) = \frac{x-x_0}{x_1-x_0} \times \frac{x-x_2}{x_1-x_2} \quad (18)$$

$$l_2(x) = \frac{x-x_0}{x_2-x_0} \times \frac{x-x_1}{x_2-x_1} \quad (19)$$

Now, calculate the $l_0(x)$, $l_1(x)$, and $l_2(x)$ for the reconstruction of the parabolic equation and retrieval of the secret key. According to Eq. 17:

$$l_0(x) = \frac{x-4}{2-4} \times \frac{x-5}{2-5} = \frac{x-4}{-2} \times \frac{x-5}{-3} = \frac{-(x^2-5x-4x+20)}{6} = \frac{-x^2-9x-20}{6}$$

According to Eq. 18:

$$l_1(x) = \frac{x-2}{2-4} \times \frac{x-5}{4-5} = \frac{x-2}{-2} \times \frac{x-5}{-1} = \frac{-(x^2-5x-2x+10)}{2} = \frac{-x^2}{2} + \frac{7x}{2} - 5$$

According to Eq. 19:

$$l_2(x) = \frac{x-2}{5-2} \times \frac{x-4}{5-4} = \frac{x-2}{3} \times \frac{x-4}{3} = \frac{x^2-4x-2x+8}{3} = \frac{x^2}{3} - 2x + \frac{8}{3}$$

Therefore, the key is constructed using Eq. 20.

$$f(x) = \sum_{j=0}^{2} y_i \times l_j(x) = y_0 l_0(x) + y_1 l_1(x) + y_2 l_2(x) \quad (20)$$

$$= 1942 \left(\frac{-x^2-9x-20}{6}\right) + 3402 \left(\frac{-x^2}{2} + \frac{7x}{2} - 5\right) + 4414 \left(\frac{x^2}{3} - 2x + \frac{8}{3}\right)$$

After solving the above Eq. 20, we get the parabolic equation:

$F(X) = 94X^2 + 166X + 1234$

Here, we can see that the Secret key = '1234' is received using the Lagrange interpolation using three different points. The secret key is used for the Involution functional-based Decryption of Multimedia File. Finally, the complete algorithm is analyzed in terms of time complexity.

**Algorithm 2: Encryption process**

```
Function LCG(seed, a, c, m, count):
   random_bits = []
   current_seed = seed

   for i from 1 to count:
      current_seed = (a× current_seed + c) % m
      random_bits.append(current_seed % 2)
   return random_bits

Function XOR(data1, data2):
   result = []
   for I  from 0 to min(length(data1), length(data2)) - 1:
result.append((data1[i] + data2[i]) % 2)
return result

Function InvolutionCipher(data, secret_key, n):
   encrypted_data = []

   for a bit of data:
      encrypted_bit = (secret_key - bit^(1/n))^n
      encrypted_data. Append(encrypted_bit)
   return encrypted_data

Function EncryptData(original_data, seed, a_lcg, c_lcg, m_lcg, count_lcg, secret_key, n):

   random_bits = LCG(seed, a_lcg, c_lcg, m_lcg, count_lcg)
   xored _data = XOR(original_data, random_bits)
   encrypted_data = InvolutionCipher(xored_data, secret_key, n)
   Output(encrypted_data)
```

**Algorithm 3:** Authorization and Key Distribution

```
Function KeyDistribution(seed, a, c, m, n):
    Output(encrypted_data)
    current_seed ← encrypted_data

Function ShamirSecretSharing(secret_key, k, n):
   coefficients ← RandomCoefficients(secret_key, k - 1)
   shares ← []

   for ← 1 to n:
      xi ← i
      yi ← EvaluatePolynomial(coefficients, xi)
      shares. Append((xi,yi))
   return shares
```

```
Function InterpolationFunction(shares):
   result ← 0

   for j ← 0 to k - 1:
      term ← shares[j].y

      form ← 0 to k - 1:
         if m ≠ j:
            term *= (x - shares[m].x) / (shares[j].x - shares[m].x)
      result += term
   return result

Function GenerateParabolicEquation(owner_attributes, secret_key):
   a0 ← secret_key
   a1 ← owner_attributes[0]
   a2 ← owner_attributes[1]

   return (a0, a1, a2)

Function AuthorizationKeyReconstruction(owner_point, receiver_point, server_point):
   k ← 3     // Minimum of three points required for reconstruction

   owner_attributes ← [owner_pointx, InterpolationFunction([owner_point, receiver_point, server_point])]
   parabolic_equation ← GenerateParabolicEquation(owner_attributes, server_point.y)
   reconstructed_key ← SolveParabolicEquation(parabolic_equation)

   return reconstructedkey

// Main Multiparty Authorization Key Generation
Function EncryptData(data, authorization_key):
   encrypted_data ← ApplyEncryptionAlgorithm(data, authorization_key)
   Output(encrypted_data)

Function MultipartyAuthorizationKeyGeneration():
   // Parameters
   seed, a, c, m ← InitializeParameters()
   n ← NumberofParticipants()

   // Key Distribution
   KeyDistribution(seed, a, c, m, n)

   //Shamir's Secret Sharing
   shares ← ShamirSecretSharing(secret_key, k, n)

   // Interpolation Function
   interpolation_result ← InterpolationFunction(shares)

   // Parabolic Equation for Key Generation
   parabolic_equation ← GenerateParabolicEquation(owner_attributes, secret_key)
```

International Journal of Computer Networks & Communications (IJCNC) Vol.17, No.6, November 2025

```
   // Authorization Key Reconstruction
   reconstructed_key  ←  AuthorizationKeyReconstruction(owner_point,  receiver_point, server_point)

   // Encryption Process
   EncryptData(original_data, reconstructed_key)
```

Table 2. Proposed Algorithm Time Complexity

| Method | Time Complexity | Description |
|---|---|---|
| Linear Congruential Generator (LCG) | $O(count\_lcg)$ | It iterates over the count to generate random bits. |
| XOR Operation | $O(min(length(N1), length(N2)))$ | It iterates over the minimum length of the two input data arrays |
| Involution Cipher | $O(length(data))$ | It iterates over each bit in the data, with simple arithmetic operations for each bit. |
| Encrypt Data Function | $O(count\_lcg + n)$ | The "EncryptData" function calls LCG, XOR, and Involution Cipher. The 'n' is the length of the original data. |
| Key Distribution Algorithm | $O(n)$ | The Involution Cipher, XOR, and Random Bits operations take $(O(1))$ time each. The loop runs (n) times, where (n) is the number of participants. |
| Shamir's Secret Sharing Algorithm | $O(n\,k)$ | Generating random coefficients involves selecting (k - 1) random values, assuming O(k) time. The loop runs (n) times for evaluating (n) shares. |
| Interpolation Function Algorithm | $O(k^2)$ | The complexity of evaluating the interpolation function depends on (k), the polynomial degree. |
| Parabolic Equation for Key Generation | $O(1)$ | It involves simple calculations. |
| Authorization Key Reconstruction Algorithm | $O(1)$ | Interpolation and solving operations are done once and do not depend on the number of participants. Solving the parabolic equation can be considered to a fixed degree. |
| Main Multiparty Authorization Key Generation | $O(n + n\,k + k^2 + 1)$ | Combining the complexities of the key distribution, Shamir's secret sharing, interpolation, and authorization key reconstruction. |

The overall time complexity of the encryption process is expressed as $O(count\_lcg + n)$, and authorization and key sharing is $O(n + nk + k^2 + 1)$. Here, the analysis considers the dominant factors and provides a high-level estimation. The specific constant factors and lower-order terms are not included in this Big O notation.

## 4. EXPERIMENTAL RESULTS AND DISCUSSION

The simulation was performed on the Windows 8.1 platform using Eclipse Indigo version 12.1. The hardware setup consisted of an Intel Core i5 processor clocked at 1.70 GHz, a single-core processor, and 4 GB of RAM running on a 64-bit operating system. Performance evaluation of multimedia files included metrics such as execution time, CPU utilization, and memory consumption. The random bitis generated using Linear Congruential Generator (LCG), and the initial parameters are as follows: Rand(•) = N1 = ($X_0$ = 9741, a = 1674, c = 1234, and m = 231) and Rep(•) = N2 = ($X_0$ = 9123, a = 1324, c = 2234, and m = 432). The random sequence was comprehensively generated in matrix format according to the file size, and the simulated outcomes were evaluated and analyzed.

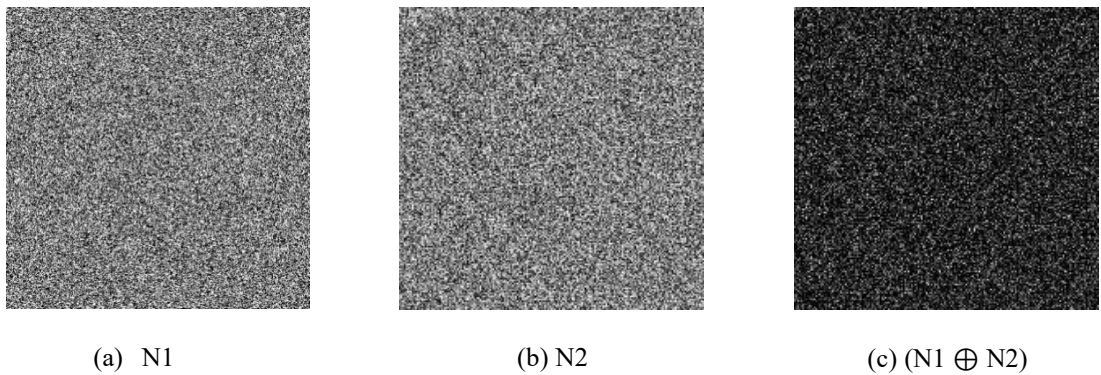

(a) N1          (b) N2          (c) (N1 ⊕ N2)

Figure 2. Random bit and XOR operation

Figure 2 illustrates how random bits are generated using a Linear Congruential Generator (LCG). The produced bits, [N1, N2], are periodically combined using XOR at intervals marked as ΔT. This XOR operation merges corresponding bits to makea new pixel density, which is then XORed with the original file. This process improves security in accordance with NIST standards and produces cryptographically secure random bits.

## 4.1. Text Encryption

To effectively handle big files and high-redundancy data, the suggested encryption technique uses an Involution Function-Based Stream Cipher. With optimal encryption speed, decryption time, file size, and throughput, it supports selective encryption. Randomly produced text files (5–30 KB) from https://www.lipsum.com/ were used for the experiments. (Sinha and others, 2020 [18]). Using the involution-based model, the system encrypts plain.txt files into ciphertext and compares its performance to that of the AES, DES, RSA, and Blowfish algorithms. To identify the most effective and secure method, metrics including file size, processing time, and throughput are examined.

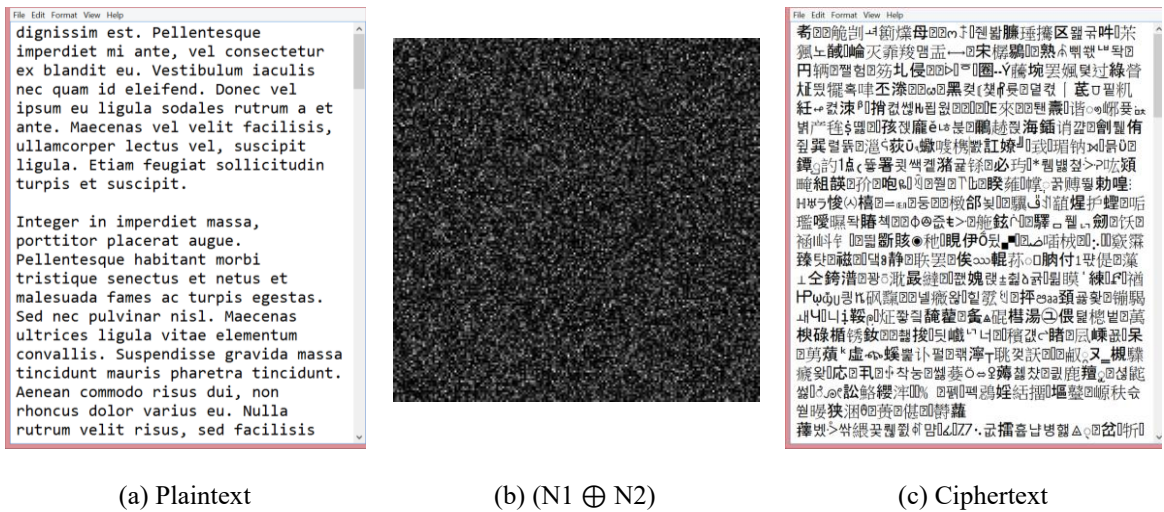

(a) Plaintext          (b) (N1 ⊕ N2)          (c) Ciphertext

Figure 3. Text Encryption and Decryption

The original text file is displayed in Figure 3(a), and the XOR random bit is shown in Figure 3(b) after being XORed with the original 5KB text file. Figure 3(c) illustrates the application of involution functional-based encryption, in which all bits are merged to create the encrypted stream cipher.

Table 3 Comparison of ciphertext with given plaintext size

| Plain Text Size in (kb) | Ciphertext Size in (kb) | | | | |
|---|---|---|---|---|---|
| | Blowfish [18] | AES [18] | DES [18] | RSA [18] | Proposed Involution Functional-based Encryption |
| 5 | 6.76 | 6.86 | 6.86 | 5.21 | 5.06 |
| 10 | 13.52 | 13.70 | 13.70 | 10.27 | 10.12 |
| 15 | 20.27 | 20.55 | 20.55 | 15.43 | 15.10 |
| 20 | 27.03 | 27.39 | 27.39 | 20.89 | 20.13 |
| 25 | 33.78 | 34.25 | 34.25 | 25.79 | 25.16 |
| 30 | 40.5 | 41.09 | 41.06 | 30.51 | 30.13 |

Table 3 presents a comparison of text files based on their size (in KB) using different encryption algorithms such as Blowfish, AES, DES, and RSA. The evaluation focuses on two important factors: (i) the size increase of the ciphertext and (ii) the encryption time. The suggested Involution Functional-Based Encryption outperforms conventional methods, resulting in fewer ciphertexts and lower storage needs, according to the results. This indicates that it is a good option for lightweight encryption applications because of its efficacy and efficiency in safe data storage.

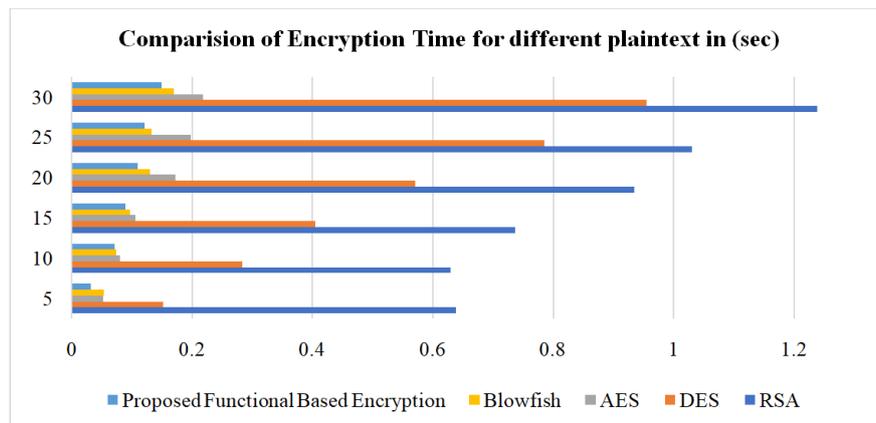

Figure 4 Comparison of encryption time (sec) for different algorithms

As shown in Figure 4, the second evaluation compares Blowfish, AES, DES, RSA, and the suggested Involution Functional-Based Encryption, with emphasis on encryption time. The suggested method achieves the fastest encryption since the involution function lowers computing complexity, whereas RSA takes longer because of the public-private key generation. Throughput, or the transmission rate of correctly encrypted data sent to distributed storage, is the third performance metric. Higher throughput values, as defined by Eq. 21, indicate improved encryption efficiency in the proposed paradigm.

$$\text{Throughput} = (\text{Plaintext size})/(\text{Encryption Time}) \quad (21)$$

Table 4. Throughput calculation for text file

| Algorithm | Blowfish [18] | AES [18] | DES [18] | RSA [18] | Proposed Involution Functional Based Encryption |
|---|---|---|---|---|---|
| Throughput (kb/sec) | 159.6 | 126.5 | 33.32 | 74.61 | 157.11 |

Table 4 displays the throughput calculations for different algorithms on text files (.txt). The throughputs for Blowfish, AES, DES, and RSA are 159.6, 126.8, 33.32, and 74.61 kb/sec, respectively [10]. Meanwhile, the proposed Involution functional-based encryption achieves a throughput of 157.11 kb/sec. These findings indicate that DES achieves the highest throughput among all traditional algorithms. In contrast, the proposed algorithm performs comparatively better than conventional techniques in terms of security[20].

## 4.2. Security and Performance Evaluation

In the distributed environment, the Cloud Service Provider (CSP) is considered semi-trusted; it follows protocols but may act for its own benefit. In contrast, the proposed system ensures trust and resilience against potential adversaries. Sensitive data resides in the cloud, while access lists are mirrored on both the organization server and the cloud. Data consumers are treated as untrusted, whereas data owners are assumed to act honestly. Although the model's strong security measures may slightly affect performance, they significantly enhance overall system reliability and robustness[20].

### 4.2.1. Attribute-based Computational Overhead

The test is constructed based on the number of attributes used for key construction.

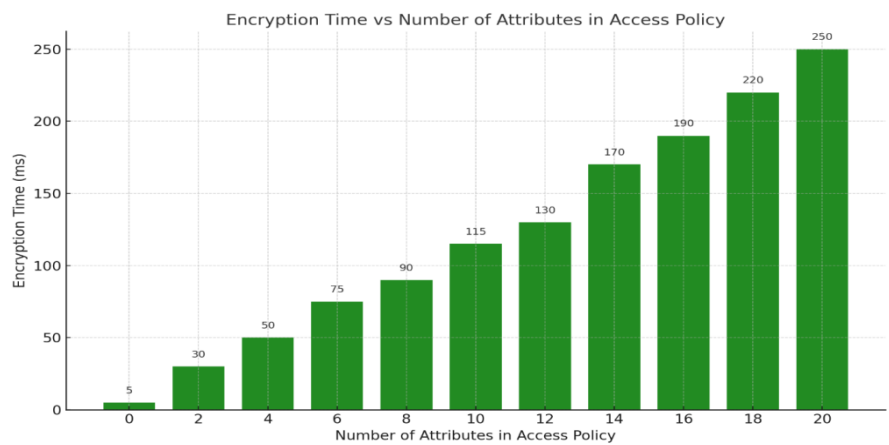

Figure 5. Encryption Time

Figure 5 represents the encryption time of the file with multiple attributes. The results show that the number of attributes in the access policy increases the time of encryption linearly.

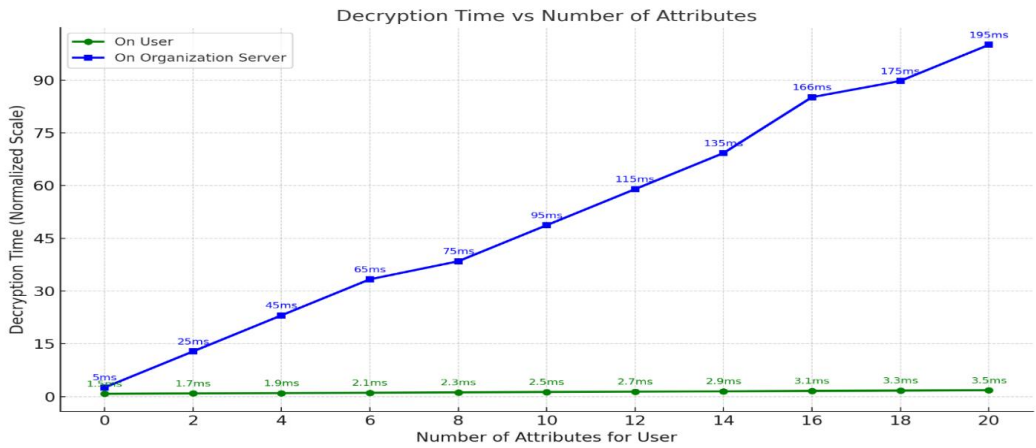

Figure 6. Decryption Time

In Figure 6, we present an analysis of the costs of decryption. The computation includes expenses associated with symmetric encryption and the distribution of authorization keys. It is important to note that when the access control module is located within the organization's storage and the backup is accessible on a cloud server, the complete attribute set is managed by multiple authorities in a collaborative effort. The cost of encryption and decryption increases linearly with the number of attributes, but execution time remains reasonable even with 20 attributes, indicating a relatively complex access policy.

**4.2.2. Storage Overhead**

In distributed, resource-constrained computing environments such as cloud platforms, edge computing, and Internet of Things-based architectures, storage overhead is a key performance metric for cryptographic access control systems. The storage needs imposed by the suggested access control model are assessed and contrasted with two benchmark schemes —DAC-MACS [12] and Sinha et al. [11] —in this section. The amount of cryptographic data that must be permanently stored at the user-side and the organization's server-side is the primary focus of the analysis.

The organization server keeps metadata for access policy enforcement, key reconstruction, and authorization tracking. At the same time, users are usually required to hold identity-based credentials and attribute-specific decryption keys in secure access control frameworks. The quantity and size of these components all directly impact system performance, scalability, and deployment viability. Table 5 presentsa detailed comparison of the storage overhead with various models.

Table 5 Storage Overhead Comparison

| Environment | DAC-MACS [12] | Sinha et al., [11] | Proposed Model |
|---|---|---|---|
| On User | $(n_{AA} + 3)\|p\|$ | $(m+1)\|Pi\|$ | $(n + 1)\|p\|$ |
| Organization Server | $(3t_c + 3)\|p\|$ | $(K_c+1)\|Pi\|$ | $(t_c + 1)\|p\|$ |

- $\|p\|$: Size of a standard cryptographic parameter (elliptic curve point, typically 256 bits).
- $\|Pi\|$: Size of a pairing-based cryptographic element (typically 512–1024 bits).
- $n_{AA}$: Number of attributes managed by the attribute authority in DAC-MACS.
- $m$: Number of attributes assigned to a user in Sinha et al.'s model.

- $n$: Number of attributes assigned to a user in the proposed model.
- $t_c$: Number of attributes embedded in the ciphertext policy.
- $K_c$: Number of key components stored per user at the server (Sinha et al. [11]).

DAC-MACS [12] imposes overhead proportional to the number of attributes from the attribute authority. Each user stores ($n_{AA}$+3) elements of size $|p|$, including versioning metadata, identity bindings, and key elements. Sinha et al. [11] reduce the number of elements stored to ($m$+1), but each element is a pairing-based component of size $|Pi|$, making them heavier in memory usage. The Proposed Model stores only ($n$+1) lightweight components of size $|P|$, where n is the number of attributes owned by the user. The model omits versioning and identity-bound key structures, significantly reducing the overall user-side memory load.

The suggested method is especially appropriate for mobile users and lightweight clients because of this optimization. Because of the intricacy of attribute versioning, key tracking, and policy mappings, DAC-MACS [12] requires the organization server to hold ($3t_c$+3) items of size $|p|$. High cumulative memory costs emerge from Sinha et al. [11] requiring the server to maintain ($K_c$+1) pairing-based components per user, where each $|Pi|$ is much larger than $|P|$. In contrast, the Proposed Model introduces an efficient design by storing only ($t_c$+1) small-sized elements. During the critical reconstruction stage, these authorization points are utilized. The related secret key can be discarded once the key has been rebuilt and safely distributed, providing additional storage relief. By reducing both the number and size of cryptographic elements stored for users and the organization's server, the system is especially suitable for real-world deployment in resource-constrained, large-scale access control environments.

### 4.2.3. Security Analysis

The security framework detailed in the preceding section is subject to a comprehensive analysis, elucidated through the following theorems:

**Theorem 1: Resilience Ensured by the Assumption of Decisional q-Parallel BDHE**

Understanding the security basis of our access control system is crucial. It relies on the decisional q-parallel BDHE assumption, which posits that adversaries cannot distinguish outputs of bilinear Diffie-Hellman exponentiation from random group elements.

**Proof:** After reconstructing the user's key on the organization's server, its structure meets the defined security framework. Confidentiality under chosen-plaintext attacks is ensured through a simulator based on the q-parallel BDHE assumption. Since solving the BDHE problem is computationally infeasible, attackers gain no advantage, confirming the system's security.

**Theorem 2: Establish that the attribute key shares in our proposed system are sufficient**

Key reconstruction requires three points: the Organization Server, Owner, and Receiver, which together generate the necessary parabolic equations. An adversary possessing fewer than j shares of a secret attribute key for any attribute obtains no meaningful information.

**Proof:** The system relies on three key points: Organization Server, Owner, and Receiver. Let Pi denote the probability that an adversary correctly guesses a secret attribute key with i shares. The Owner Point serves to verify the secret and prevent unauthorized access. According to Theorem 2, security holds if the attribute set meets the encryption policy and the data owner authorizes access with enough attribute key shares.

146

**Theorem 3: Resistance to Collusion Attacks**

The proposed system prevents collusion attacks by assigning each user a unique combination of *User_Id (Ui)*, *User_Type (UTi)*, and *User_Credentials (UCi)*, embedded within a parabolic equation linked to secret keys. Access to encrypted data requires both the Organization Server Point and the Owner Point, allowing only one receiver at a time. This design ensures that multiple users cannot combine credentials to decrypt the ciphertext, effectively blocking unauthorized access.

**Proof:** Every data consumer gets a distinct group of credentials: User_Id (Ui), User_Type (UTi), and User_Credentials (UCi), which are combined in a parabolic equation by Involution Function-Based Encryption. The access requires a conjunction of the Organization Server Point and the Owner Point. Hence, only one receiver can decrypt data at a time. This method mathematically forbids the simultaneous decryption of data by multiple users and, therefore, is very effective at protecting against collusion attacks and unauthorized access.

## 5. CONCLUSION

This research uses Lagrange interpolation and Ciphertext-Policy Attribute-Based Encryption (CP-ABE) to propose a novel access-control framework for cloud computing. The method addresses issues with conventional, inflexible policy models that make key management and storage difficult by providing both complete security and user-friendly administration. By reconstructing secret keys using Shamir's Secret Sharing and a threshold-based multiparty protocol, data leakage is reduced and strict data ownership is enforced. By managing secret shares and releasing decryption keys only when specific access policies are met, the organization's server plays a semi-trusted role, reducing the risk of collusion, unauthorized access, and user revocation issues. It is challenging for unauthorized users, even when working together, to rebuild keys because they are closely tied to user attributes. To enhance trust and traceability, the framework also offers advanced features such as blockchain-based audits, customized access controls, and reputation-aware re-encryption. These features improve secure data management across a variety of industries, including cloud services, healthcare, the military, and cooperative data-sharing platforms.

## CONFLICT OF INTEREST

The authors declare no conflict of interest.

**AUTHORS**

**Dr. Keshav Sinha** is an Assistant professor specializing in cryptography and network security, with research focused on developing secure environments for multimedia transmission. His contributions extend beyond cryptography into Soft Computing, Humanities, and cyber policy, emphasizing innovation and flexibility in computer science. With a strong academic background and numerous publications in reputable conferences and journals, he has established himself as a dedicated researcher and educator. His expertise in cryptography and cybersecurity, coupled with his interdisciplinary approach, has positioned him as a thought leader in the field. His work in emerging security frameworks, regulatory challenges, and AI-driven cryptographic advancements continues to shape the future of digital security.

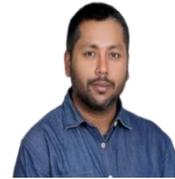

**Dr. Sumitra** is an Assistant Professor at UPES, Dehradun, India. She earned her Ph.D. from the Birla Institute of Technology and Science (BITS) Pilani, Pilani Campus. Her research primarily focuses on IoT security, artificial intelligence and machine learning (AI/ML), and federated learning. Dr. Sumitra has contributed extensively to the field through publications in reputed international journals, including the Journal of Network and Computer Applications and Internet of Things. Her research excellence was recognized with the Outstanding Research Article Award at the Doctoral Colloquium, BITS Pilani, for her work titled "RPL: An Explainable AI-based Routing Protocol for the Internet of Mobile Things."

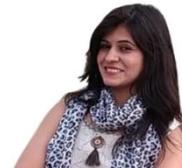

**Dr. Richa Kumari** is serving as an Assistant Professor at UPES, Uttarakhand. She holds a Ph.D. in Computer Science and Engineering from MNIT Jaipur. Her research expertise spans Device-to-Device (D2D) communication, federated learning, wireless networking, IoT, and resource allocation in B5G networks. She has published extensively in reputable venues and has qualified for both the GATE and the UGC-NET. Earlier, she worked as a Junior Research Fellow in a DST Rajasthan-funded project. She also served as Faculty at IIIT Kota and the Government Women's Engineering College, Ajmer. Additionally, she gained teaching experience as a Teaching Assistant at MNIT Jaipur during her PhD. She has actively contributed as a reviewer for various conferences and research forums.

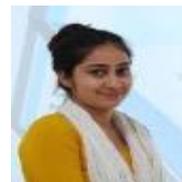

**Dr. Akashdeep Bhardwaj** is a Professor and Head of the Cybersecurity Center of Excellence at UPES, Dehradun, India and CEO for Global Cybersecurity Association (GCA). An eminent IT Industry expert in Cybersecurity, Digital Forensics, and IT Operations, Dr. Akashdeep mentors graduate, master's, and doctoral students and leads several projects. Dr. Akashdeep is a Postdoctoral Fellow from Majmaah University, Saudi Arabia, with a Ph.D. in Computer Science, a Postgraduate Diploma in Management (equivalent to an MBA), and an Engineering Degree in Computer Science. Dr. Akashdeep has published 180 research works (including copyrights, patents, papers, authored & edited books) in international journals.

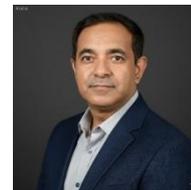

**Dr. Shawon Rahman** is a tenured professor of Computer Science and Engineering at the University of Hawaii-Hilo, a Faculty Applied Clean Energy Sciences (FACES) Fellow at the National Renewable Energy Laboratory, and an adjunct faculty member at the University of Missouri–Kansas City. With more than 19 years of teaching experience, he has chaired and supervised numerous Ph.D. dissertations. Dr. Rahman has secured and managed several federal, state, and foundation grants, including those from NSF, USDA, and DOE. His research spans Cybersecurity, Digital Forensics, Cloud Computing, and STEM outreach. A senior member of IEEE, he has published over 140 peer-reviewed papers and actively serves on editorial boards, professional committees, and national review panels.

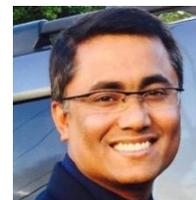